\documentclass[aps,twocolumn,floatfix,altaffilletter,superscriptaddress,preprintnumbers,
               tightenlines,showpacs,showkeys,notitlepage,nofootinbib]{revtex4-2}

\usepackage{amsmath,amssymb} 
\usepackage{graphicx} 
\usepackage[export]{adjustbox} 
\usepackage{url} 
\usepackage[dvipsnames]{xcolor} 
\usepackage{siunitx} 
\usepackage{etoolbox,orcidlink} 
\usepackage{tabularx,booktabs} 
\usepackage{microtype}
\usepackage[normalem]{ulem} 

\makeatletter
\def\thesection{\arabic{section}}%
\def\p@section{}%
\def\thesubsection{\thesection.\arabic{subsection}}%
\def\p@subsection{}%
\def\thesubsubsection{\thesubsection.\arabic{subsubsection}}%
\def\p@subsubsection{}%
\def\appendix{%
    \par
    \setcounter{section}\z@
    \setcounter{subsection}\z@
    \setcounter{subsubsection}\z@
    \def\thesubsection{\thesection.\arabic{subsection}}%
    \def\thesubsubsection{\thesubsection.\arabic{subsubsection}}%
    \def\p@subsection{}%
    \def\p@subsubsection{}%
    \@addtoreset{equation}{section}%
    \def\theequation@prefix{\thesection}%
    \addtocontents{toc}{\protect\appendix}%
    \@ifstar{%
    \def\thesection{\unskip}%
    \def\theequation@prefix{A.}%
    }{%
    \def\thesection{\Alph{section}}%
    }%
}%
\makeatother

\def\apjl{Astrophys.~J.~Lett.}

\def\apjs{Astrophys.~J.~Suppl.}

\usepackage{hyperref} 
\usepackage[capitalize]{cleveref} 
\newcommand\myshade{80} 
\colorlet{mylinkcolor}{ForestGreen}
\colorlet{mycitecolor}{Red}
\colorlet{myurlcolor}{violet}
\hypersetup{
  linkcolor  = mylinkcolor!\myshade!black,
  citecolor  = mycitecolor!\myshade!black,
  urlcolor   = myurlcolor!\myshade!black,
  colorlinks = true
}

\newcommand{\ev}[1]{\ensuremath{\left\langle #1 %
                     \right\rangle}} 

\newcommand{\rbubble}{\ensuremath{r_*}}   

\newcommand{\github}[1]{\href{https://github.com/#1}{\includegraphics[width=8pt]{Plots/github.pdf}}}

\makeatletter
\providecommand*{\diff}%
	{\@ifnextchar^{\DIfF}{\DIfF^{}}}
\def\DIfF^#1{%
	\mathop{\mathrm{\mathstrut d}}%
		\nolimits^{#1}\gobblespace}
\def\gobblespace{%
	\futurelet\diffarg\opspace}
\def\opspace{%
	\let\DiffSpace\!%
	\ifx\diffarg(%
		\let\DiffSpace\relax
	\else
		\ifx\diffarg[%
			\let\DiffSpace\relax
		\else
  			\ifx\diffarg\{%
				\let\DiffSpace\relax
			\fi\fi\fi\DiffSpace}

\begin{document}

\title{High-Frequency Gravitational Waves from Phase Transitions in Nascent Neutron Stars}

\author{Katarina Bleau \orcidlink{0009-0005-5380-5253}}
\email{kbleau@uni-mainz.de}
\affiliation{PRISMA Cluster of Excellence \& Mainz Institute for
             Theoretical Physics, \\
             Johannes Gutenberg University, 55099 Mainz, Germany}

\author{Joachim Kopp \orcidlink{0000-0003-0600-4996}\,}
\email{jkopp@cern.ch}
\affiliation{PRISMA Cluster of Excellence \& Mainz Institute for
             Theoretical Physics, \\
             Johannes Gutenberg University, 55099 Mainz, Germany}
\affiliation{Theoretical Physics Department, CERN,
             1211 Geneva 23, Switzerland}

\author{Jiheon Lee \orcidlink{0000-0001-7235-5423}}
\email{anffl0101@kaist.ac.kr}
\affiliation{Department of Physics, Korea Advanced Institute of
             Science and Technology, Daejeon 34141, Korea}
\affiliation{Theoretical Physics Department, CERN,
             1211 Geneva 23, Switzerland}

\author{Jorinde van de Vis \orcidlink{0000-0002-8110-1983}}
\email{jorinde.van.de.vis@cern.ch}
\affiliation{Theoretical Physics Department, CERN,
             1211 Geneva 23, Switzerland}

\date{\today}

\preprint{CERN-TH-2026-052, MITP-26-010} 

\begin{abstract}
\noindent
Tentative evidence suggests that the cores of massive neutron stars consist of deconfined quark matter. We argue that the formation of such a quark matter core during a galactic supernova could be accompanied by the emission of gravitational waves in the MHz band. These signals constitute a new target for high-frequency gravitational wave detectors, demonstrating that such detectors may offer unique opportunities for testing quantum chromodynamics in an otherwise inaccessible regime.
\end{abstract}

\maketitle

Neutron stars are unique laboratories for particle physics under extreme conditions. In particular, they probe quantum chromodynamics (QCD) at very large particle density (or chemical potential, $\mu$) and relatively low temperature, a region of the QCD phase diagram that is otherwise all but inaccessible both experimentally and theoretically. At very large $\mu$, hadronic matter is believed to undergo a first-order phase transition to a plasma of free quarks and gluons \cite{%
  Shuryak:1980tp,   
  Stephanov:2004wx, 
  Rajagopal:2000wf, 
  Fukushima:2010bq, 
  Shuryak:2014zxa,  
  ALICE:2022wpn}.   
However, the exact location of the phase boundaries in the temperature ($T$) vs.\ $\mu$ plane is unknown.

Recently, evidence has been accumulating that in the cores of sufficiently massive neutron stars, conditions are indeed such that matter exists in the form of deconfined quarks 
 \cite{Annala:2019puf,   
       Annala:2023cwx,   
       Ayriyan:2025rub}. 
This implies that during the formation of a neutron star in a supernova, its core may transition to a quark phase.

Here, we argue that, if the transition proceeds via the nucleation, expansion, and coalescence of quark matter bubbles, it could source \emph{high-frequency gravitational waves} (GWs) at $\gtrsim \si{MHz}$ frequencies from the acoustic reverberations in the stellar core caused by bubble collisions. Our results imply that high-frequency GW detectors, whose physics program has so far revolved around probing exotic physics like primordial black holes, superradiance, or signals from the very early Universe \cite{Aggarwal:2025noe}, may offer a unique and novel way of probing the QCD phase transition at large $\mu$. The existence of the signal as well as its amplitude depend strongly on the neutron star equation of state. Moreover, any signals will be rare: the supernova rate in the Milky Way is only 1--2 per century \cite{Rozwadowska:2020nab}.  But similar to other rare astrophysical signals like supernova neutrinos, the potential payoff is well worth taking chances and waiting.

GW production in first-order phase transitions has been studied in detail in the context of the early Universe \cite{
  Witten:1984rs,     
  Caprini:2009yp,    
  Durrer:2010xc,     
  Hindmarsh:2013xza, 
  Schwaller:2015tja, 
  Breitbach:2018ddu, 
  Cutting:2020nla,   
  Hindmarsh:2020hop, 
  LISACosmologyWorkingGroup:2022jok,  
  Correia:2025qif,   
  Caprini:2024gyk}.  
Their emission due to QCD dynamics in a neutron star has been discussed in ref.~\cite{Blas:2022xco}, focusing on phase transitions triggered by a binary neutron star merger.
The effect of a phase transition on the GW signal from the collapse and the postmerger evolution has been studied in refs.~\cite{Yasutake:2007st, Dexheimer:2018dlz, Most:2018eaw, Bauswein:2018bma, Zha:2020gjw, Kuroda:2021eiv}.

In the following, we first outline the dynamics of the QCD phase transition under supernova conditions and we discuss the associated uncertainties. We then explain how we model the neutron star, in particular which equations of state (EoS) we consider. We finally estimate the frequency and amplitude of the high-frequency GWs emitted during a first-order QCD phase transition associated with a supernova, and we discuss the detectability of such a signal. In the appendices, we provide supplementary technical details, and we comment on stochastic signals from past supernovae across the Universe.


\textbf{Phase transitions in collapsing stars.} When the core of a massive star collapses under its own gravity, initiating a supernova and forming a neutron star, the pressure at its center may exceed first the critical pressure, $p_c$, beyond which deconfined quark matter is thermodynamically more stable than hadronic matter, and then the nucleation pressure, $p_n$, at which the first quark matter bubble actually forms. The ensuing transition could proceed in several different ways:
\begin{enumerate}
    \setlength{\itemsep}{3pt}
    \setlength{\parskip}{0pt}
    \item A complete transition, where quark matter bubbles expand and coalesce until the whole core has been converted to the quark phase. A complete transition would yield the strongest GW signal, sourced mostly by sound waves. However, a complete transition can only occur if
    \begin{align}
        n_{b,h}(p_n) > n_{b,q}(p_c) ,
        \label{eq:nB-condition}
    \end{align}
    where $n_{b,h}(p_n)$ ($n_{b,q}(p_c)$) is the baryon number density in the hadron (quark) phase at the nucleation pressure (at the critical pressure). 
    For the equations of state (EoS) considered in this work, no stable neutron stars exist with core pressures satisfying \cref{eq:nB-condition}.

    \item A stalled transition, where \cref{eq:nB-condition} is not satisfied and the expansion of quark matter bubbles stops before the transition is complete, forming a (macroscopically) \emph{mixed phase}. This is the situation we find for all equations of state considered in this work. For a stalled transition, we still expect a GW signal, but it will be suppressed compared to a complete transition.

    \item A transition involving a microscopically mixed phase \cite{Glendenning:1992vb, Glendenning:2001pe, Mintz:2009ay}. In this scenario (``Gibbs construction'') quarks and hadrons co-exist, either as an unordered mixture of two liquids, or in the form of ``pasta'', where quark matter forms regular structures like nodules (``gnocchi''), layers (``lasagna''), or rods (``spaghetti''). The volume fraction covered by these structures then gradually increases. The formation and expansion of a microscopically mixed phase is not expected to lead to GW emission.
\end{enumerate}
Whether the phase transition occurs via bubbles or via microscopic structure depends on the surface tension \cite{Glendenning:1992vb, Alford:2001zr}.
The high-frequency GW signal potentially accompanying a supernova thus depends very strongly on the detailed dynamics of QCD at extreme baryon number density. 
Conversely, this means that searches for such a signal coincident with the next galactic supernova can provide crucial information about a poorly understood facet of the Standard Model of particle physics that we cannot access in any other way.


\begin{figure}
    \begin{tabular}{c@{\ \ }c}
        \hspace{-0.2cm}\includegraphics[width=0.5\linewidth]{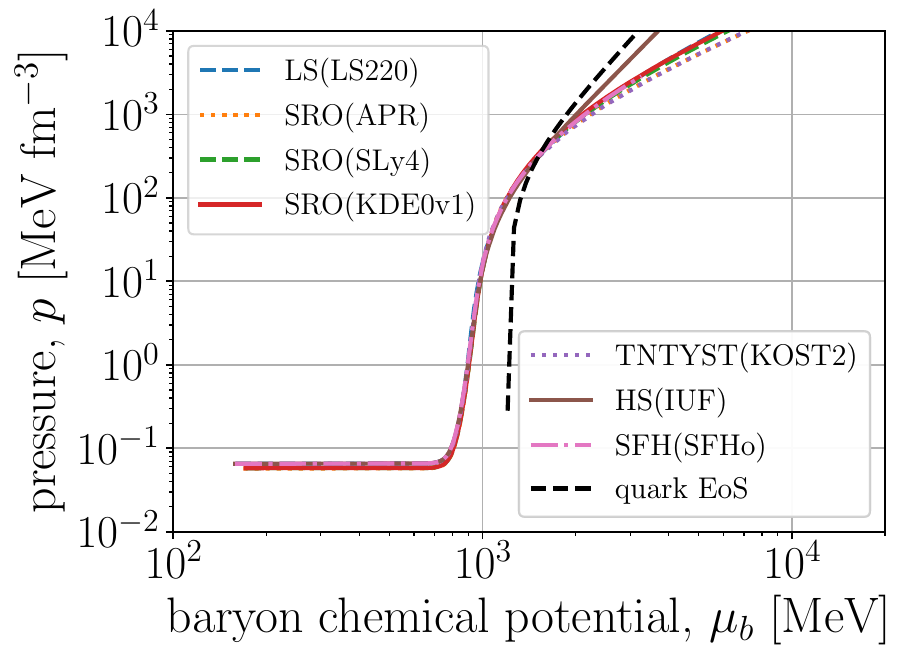} &
        \raisebox{-0.0cm}{\includegraphics[width=0.5\linewidth]{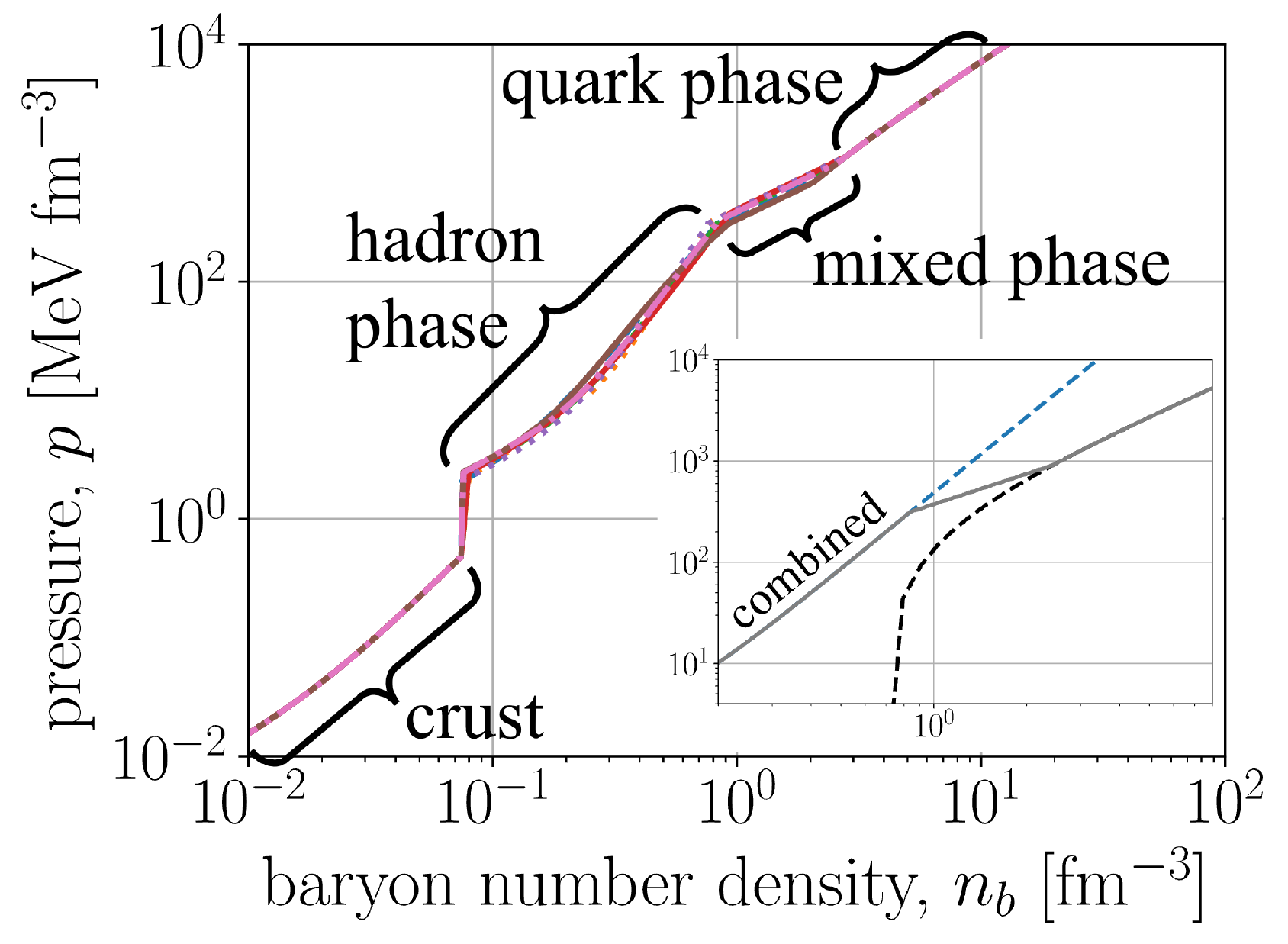}} \\
        (a) & (b)
    \end{tabular}
    \caption{Neutron star equations of state for both hadronic matter and quark matter.  For the latter, we use a bag equation of state. In the $p$-vs.-$\mu_b$ plane (\emph{left panel}), the crossing point between the hadron (coloured curves) and quark (black dashed curve) EoS defines the critical pressure. In the $p$-vs.-$n_b$ plane (\emph{right panel}), the various phases can be clearly distinguished. The hadron and quark EoS are plotted separately in the inset.}
    \label{fig:eos}
\end{figure}

\textbf{Modelling a neutron star.}
To describe a neutron star with a quark matter core, we need a pair of equations of state, one for the hadron phase and one for the quark phase. We choose several hadron EoS \cite{Raduta:2021coc, LATTIMER1991331, Schneider:2019vdm, Schneider:2017tfi, Togashi:2017mjp, Hempel:2009mc, Steiner:2012rk} from the CompOSE database \cite{Typel:2013rza, Oertel:2016bki, CompOSECoreTeam:2022ddl}, while for the quark phase, we use a simple bag EoS \cite{Alford:2004pf} of the form
\begin{align}
    p = -C + \frac{3}{4\pi^2} a_4 \mu_q^4 \,,
    \qquad
    e = 4C + 3p,
    \label{eq:bag}
\end{align}
with $C = (200$~MeV)$^4$ and $a_4 = 0.81$, where the quark and baryon chemical potentials are related as $\mu_q = \mu_b/3$. All EoS used in this paper are plotted in \cref{fig:eos}~(a) at a typical supernova core temperature $T = \SI{30}{MeV}$, and for a charge fraction consistent with beta equilibrium.
The point where the quark and hadron EoS cross defines the critical pressure, $p_c$.\footnote{Equal $p$ on both sides of the phase boundary ensures mechanical equilibrium, while equal $\mu_b$ ensures chemical equilibrium.
}

In the $p$-vs.-$n_b$ plane (\cref{fig:eos}~(b)), a phase transition manifests as a discontinuity in $n_b$ at the critical pressure. If a mixed phase exists, as is the case for all EoS considered here, the discontinuity is smoothed out. We parameterize the mixed phase with a polytrope of the form $p = \kappa n_b^{\Gamma}$ \cite{Annala:2019puf}, with its starting point being $n_{b,h}(p_c)$ and its endpoint being its intersection with the quark EoS.

After bubble nucleation has started at $p_n$, hadronic matter will convert to quark matter until the chemical potential has dropped back to $\mu_{b,c}$, which is the point at which further conversion becomes thermodynamically unfavourable. We denote the corresponding pressure with $p_s$ ($s$ for `stall'). For a given choice of $p_s$, we can determine the fraction $x_q$ of the volume that transitions before bubble expansion stalls from the conservation of baryon number density between the two phases:
\begin{align}
    x_q n_{b,q}(\mu_{b,c}) + (1-x_q) n_{b,h}(\mu_{b,c}) = n_{b,h}(\mu_{b,s}).
    \label{eq:xq-from-eos}
\end{align}

\begin{figure}
    \centering
    \includegraphics[width=0.95\linewidth]{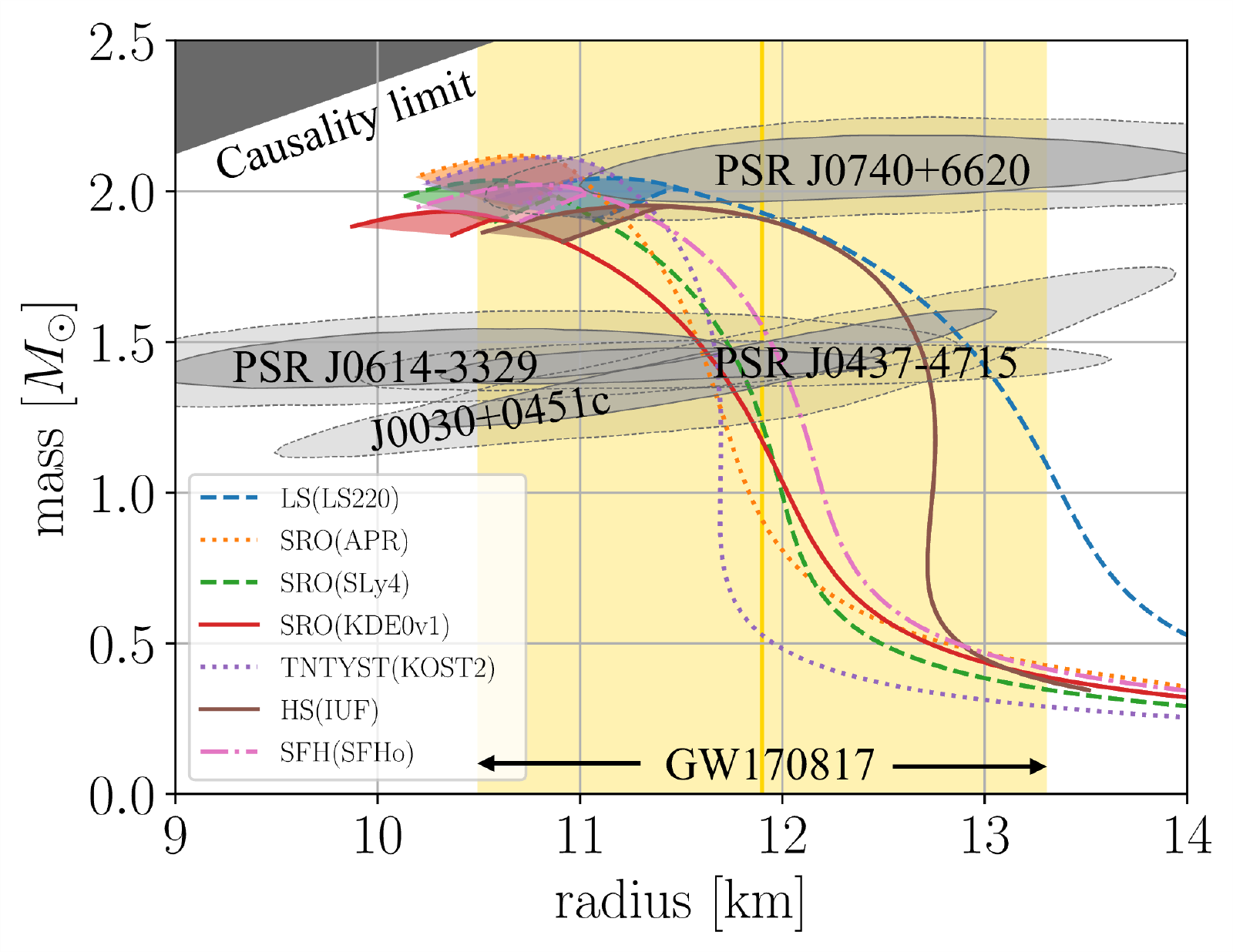}
    \vspace*{-1.5em}
    \caption{Mass--radius relations for the neutron star equations of state considered in this work (coloured curves), compared to constraints from NICER (light grey) \cite{
      Mauviard:2025dmd,    
      Salmi:2024aum,       
      Vinciguerra:2023qxq, 
      Choudhury:2024xbk,   
      Shirke:2025gfi},     
      GW170817 (yellow) \cite{LIGOScientific:2018cki}, and the theoretical limit from causality (dark grey) \cite{Demorest:2010bx}. At large masses, the curves fan out to reflect the variation due to the choice of the exponent $\Gamma$ ($0.3 \leq \Gamma \leq 1.2$) in the polytrope describing the mixed phase.}
    \label{fig:m-r}
\end{figure}

For each combined EoS, we solve the Tolman--Oppenheimer--Volkoff (TOV) equations to relate the mass, radius, and core pressure of the star in equilibrium. The solution to the TOV equations also determines the radius of the quark matter core $R_c$, characterized by the condition $p > p_c$. Values of $R_c$ for the EoS considered in this paper are listed in the second column of \cref{tab:params}, assuming in each case the maximum stable neutron star mass.
The TOV equations also yield the mass--radius relation, which we plot (for $T = 0$) in \cref{fig:m-r} together with constraints on neutron star masses and radii from NICER \cite{Mauviard:2025dmd, Salmi:2024aum, Vinciguerra:2023qxq,  Choudhury:2024xbk, Shirke:2025gfi}, and LIGO/Virgo \cite{LIGOScientific:2018cki}.

\begin{table*}[]
    \begin{tabular*}{\linewidth}{@{\extracolsep{\fill}} lccccc}
        \toprule
        Hadron EoS   & $R_c$ [km] & $v_w$ & $x_q$ & $\sigma$ [MeV/fm$^2$]& $f^{\rm peak}$ [MHz] \\
        \midrule
        LS(LS220) \cite{Lattimer:1991nc, Oertel:2012qd}
                     & 0.75       & 0.066                 & 0.0075      & 8.01           & 1.6                   \\
        SRO(APR) \cite{Akmal:1998cf, Schneider:2017tfi, Schneider:2019vdm}
                     & 0.46       & 0.038                 & 0.0025      & 5.24          & 583                   \\
        SRO(SLy4) \cite{Schneider:2017tfi, Lattimer:1991nc, Chabanat:1997un}    
                     & 0.62       & 0.052                 & 0.0050      & 6.38          & 2.3                   \\
        SRO(KDE0v1) \cite{Lattimer:1991nc, Agrawal:2005ix, Schneider:2017tfi}
                     & 2.82       & 0.23                 & 0.15      &  96.1         & 0.16                   \\
        TNTYST(KOST2) \cite{Togashi:2017mjp}
                     & 0.061       & 0.031                 & 0.0020      & 3.96          & 498                   \\
        HS(IUF) \cite{Hempel:2009mc, Fattoyev:2010mx, Roca-Maza:2008hmq, Hempel:2011mk, Steiner:2012rk}
                     & 1.36       & 0.15                 & 0.061     & 17.9           & 1.5                   \\
        SFH(SFHo) \cite{Hempel:2009mc, Moller:1996uf, Steiner:2012rk, Hempel:2011mk}
                     & 0.81       & 0.066                 & 0.0082     & 8.86           & 1.5                   \\
        \bottomrule
    \end{tabular*}
    \caption{Phase transition parameters underlying the characteristic strain calculation in \cref{fig:h-c,fig:gw-spectra} for the different hadron EoS considered in this paper. The columns from left to right are the radius of the quark matter core (the region where $p > p_c$), the bubble wall velocity, the final quark matter fraction (see \cref{eq:xq-from-eos}), the bubble surface tension, and the peak GW frequency. Every quantity is calculated at the stalling pressure where the expected bubble number decreases to 2 for the first time. For HS(IUF) and SRO(KDE0v1), we use instead the maximum pressure before the star becomes unstable as $\geq 2$ bubbles are expected even at that pressure. This choice corresponds to the largest likely GW signal.}
    \label{tab:params}
\end{table*}

\textbf{Gravitational wave frequency.} The frequency spectrum of GWs produced during a first-order QCD phase transition in the Maxwell construction (i.e.\ in absence of a microscopically mixed phase) peaks at wavelengths comparable to the typical size $\rbubble$ of the quark matter bubbles when they collide. To estimate the peak frequency, our starting point is the bubble nucleation rate per unit volume \cite{Coleman:1977py, Callan:1977pt, Mazumdar:2018dfl, Caprini:2019egz},
\begin{align}
    \frac{dP_n(t)}{dt\,d^3x} = \Lambda^4 e^{-S(t)} \,,
\end{align}
where $S(t)$ is the bounce action at time $t$ and $\Lambda$ is an energy scale characteristic of the phase transition. In the following, we will use $\Lambda = \SI{200}{MeV}$, but we will see that our results do not depend on $\Lambda$ independently, but only on a combination of $\Lambda$ and the bubble surface tension $\sigma$, which we compute below.

After the neutron star reaches, at time $t_c$, the critical density beyond which quark matter is more stable than hadronic matter, the probability of nucleating a bubble in an infinitesimal time interval $dt'$ is
\begin{align}
    dP_n = \frac{4}{3} \pi R_c^3 \Lambda^4 e^{-S(t')} dt' \,.
\end{align}
The fraction of the core that has transitioned to the quark phase at time $t > t_c$ is then approximately given by \cite{Enqvist:1991xw}
\begin{align}
    x_q(t) = 1 - \exp\bigg[-
        \int_{t_c}^t \!dt' \tfrac{4}{3} \pi v_w^3 (t-t')^3 \Lambda^4 e^{-S(t')}
    \bigg]\,,
    \label{eq:f-true}
\end{align}
where $v_w$ is the bubble expansion velocity. Given an equation of state, we obtain $v_w$ by assuming local thermal equilibrium (LTE) and solving the hydrodynamic equations for pure radiation, see \cref{sec:fluid-dynamics}.
The number density of quark matter bubbles at time $t$ is then
\begin{align}
    n_{\rm bubbles}(t) = \int_{t_c}^t \! dt' \, \big[ 1 - x_q(t') \big]
    \frac{dP_{\rm nuc}(t')}{dt'\,d^3x}\,,
    \label{eq:n-bubbles}
\end{align}
and the peak GW frequency can be estimated as
\begin{align}
    f^{\rm peak}(t) \simeq \langle \rbubble(t) \rangle^{-1}
                   &= [n_{\rm bubbles}(t) / x_q]^{1/3} \,,
    \label{eq:f-gw}
\end{align}
where $\langle \rbubble(t) \rangle$ is the average bubble radius.

To evaluate \cref{eq:f-gw}, we parameterizate the action using the thin-wall approximation:
\begin{align}
    S(t) \simeq \frac{27 \pi^2}{2} \frac{\sigma^4}{(\Delta V(t))^3} \,,
\end{align}
where $\sigma$ is the surface tension of the bubbles and $\Delta V(t)$ is the free energy (= pressure) difference between the hadron and quark phases at time $t$. $\sigma$ is yet to be determined, while $\Delta V(t)$ can be obtained from the equation of state, provided we can relate $t$ to the fluid properties, in particular $p$. We derive $p(t)$ based on the relation between the core pressure and the neutron star mass, $M$, from the TOV equations, and on the mass growth function of the fast-accreting nascent neutron star, $M(t)$, taken from supernova simulations \cite{Suwa:2014sqa, Shankar:2025pwq}. $p(t)$ is an important source of uncertainty in our calculation. We then solve \cref{eq:f-true} at the time $t_s$ when the phase transition stalls to obtain $\sigma$:
\begin{multline}
    x_q(t_s) = 1 - \exp\bigg[
        -\int_{t_c}^{t_s} \! dt' \, \Lambda^4 e^{-S(t')} \\
            \times \frac{4}{3} \pi \min\big[ x_q(t_s) R_c^3, \; v_w^3 (t_s - t')^3\big] 
      \bigg] \,.
    \label{eq:xq-at-ts}
\end{multline}
Here, $x_q(t_s)$ is calculated according to \cref{eq:xq-from-eos}. Note that in the integrand we restrict bubbles to remain smaller than the total volume that converts to the quark phase. It is now evident why $\Lambda$ is not an independent parameter: any change in $\Lambda$ would be absorbed into a corresponding shift in $\sigma$. Note that numerical accuracy is important here: $\sigma$ appears in the exponent of an exponent, so a small error in $\sigma$ implies an enormous deviation from the desired $x_q(t_s)$.

Lacking a full microscopic theory, we treat $t_s$ (or, equivalently, $p_s$) as a free parameter. When $p_s$ is close to $p_c$, many bubbles can nucleate quickly, but $x_q(t_s)$ is small, so the probability that bubbles collide before stalling is low. Conversely, at large $p_s$, only a single bubble may nucleate. In both cases, GW emission would be strongly quenched or completely absent. We display our estimates for $f^{\rm peak}$ in the rightmost column of \ref{tab:params}, where for illustrative purposes, we have set $p_s$ to the pressure at which on average two quark matter bubbles are expected to form before the transition stalls. We see that typical GW frequencies are $\gtrsim \SI{1}{MHz}$.


\textbf{Gravitational wave amplitude.} 
We now describe how we estimate the amplitude of GWs from a QCD phase transition inside a neutron star. This discussion applies to complete phase transitions as well as stalled transitions with a macroscopically mixed phase.  As argued above, for phase transitions involving a microscopically mixed phase (such as ``pasta''), we do not expect significant GW emission.
Following refs.~\cite{%
  Maggiore:2018sht,             
  Moore:2014lga,
  Blas:2022xco},
we parameterize the observed characteristic strain as
\begin{align}
    (h_c^{\rm obs})^2 = \frac{d\big\langle(h^{\rm obs})^2\big\rangle}{d\log f}
        &= \frac{\tau^2}{d^2} \frac{d\langle h^2\rangle}{d\log f}
         = \frac{\tau^2}{d^2}
          \frac{8\pi G}{2\pi^2 f^2}
                \frac{d\rho_{\rm gw}}{d\log f} ,
    \label{eq:hc}
\end{align}
where $d$ is the distance to the neutron star; $\tau$ the sourcing time, i.e.\ the duration of the GW emission; $G$ Newton's constant; $f$ the GW frequency; and $\rho_{\rm gw}$ the GW energy density in the source at the end of the phase transition.
The factor $\tau^2/d^2$ accounts for the dilution of the GW energy density during propagation: energy that is originally contained in a sphere of radius $\tau$ dilutes into a spherical shell of radius $d$ and thickness $\tau$. (We have dropped $\mathcal{O}(1)$ factors here.)

We write the energy density as \cite{Hindmarsh:2015qta, Blas:2022xco}
\begin{align}
    \frac{d\rho_{\rm gw}}{d\log f} = 8 \pi G \rho_{\rm kin}^2 \,
        \tau \rbubble \,
        \frac{4\pi f^3}{c^3} \rbubble^3 \,
        \frac{R_c^3}{\tau^3}
        \, \tilde P_{\rm gw}(z) \, x_q^{4.8},
    \label{eq:rho-gw}
\end{align}
with $\rho_{\rm kin}$ the kinetic energy density in the fluid, determined from the hydrodynamics solution. The dimensionless spectral density, $\tilde P_{\rm gw}(z)$, is a function of $z = k \rbubble$, with $k$ the wave number, and encodes the dynamics of GW; we compute it using the sound shell model \cite{Hindmarsh:2016lnk, Hindmarsh:2019phv}, with the fluid velocity profiles as input, see \cref{sec:fluid-dynamics} for details.

The empirical factor $x_q^{4.8}$ accounts for the suppression of GW emission for incomplete ($x_q < 1$) phase transitions. To obtain this scaling, we have carried out numerical simulations of bubble nucleation, expansion, and percolation. On a periodic lattice with $80^3$ sites, we randomly nucleate bubbles following a probability distribution $P^{\rm nuc}(t) \propto e^{\beta t}$, where $\beta$ is a parameter which we choose to be $\mathcal{O}(1)$. Once nucleated, bubbles expand with constant wall velocity. Each bubble wall segment carries along a fluid perturbation whose energy density grows with the surface area of the wall segment and with the bubble radius. When wall segments collide, their associated fluid energy is added to the free sound wave energy, which ultimately determines the GW energy density. More details are given in \cref{sec:xq-scaling}. Using this method, we estimate that $\rho_\text{gw} \propto x_q^{4.8}$. Note that \cref{eq:rho-gw} also depends on $x_q$ through $\tau$ and $\rbubble$. 

The factor $R_c^3 / \tau^3$ in \cref{eq:rho-gw} accounts for differences between GW production from phase transitions in the early Universe (the case for which \cref{eq:rho-gw} has been derived and numerically verified) compared to phase transitions inside a neutron star. Namely, in the latter case the sourcing occurs in a finite volume, whereas in the early Universe, it happens everywhere.

We identify the sourcing time of GWs with the shock formation time $\tau_{\rm sh} = \ev{\rbubble} / \bar{U}_f$ \cite{Caprini:2019egz, Hindmarsh:2015qta}, where $\bar{U}_f$ is the root mean square fluid velocity.
$\tau_{\rm sh}$ is the time after which the formation of dissipative acoustic shocks ends sound wave-induced GW generation \cite{Caprini:2019egz}.
We find that $\tau_{\rm sh}$ is $\sim 10$--100\,\si{\micro sec}, which is about 1--2 orders of magnitude larger than the duration of the transition.


\begin{figure}
    \centering
    \includegraphics[width=0.9\columnwidth]{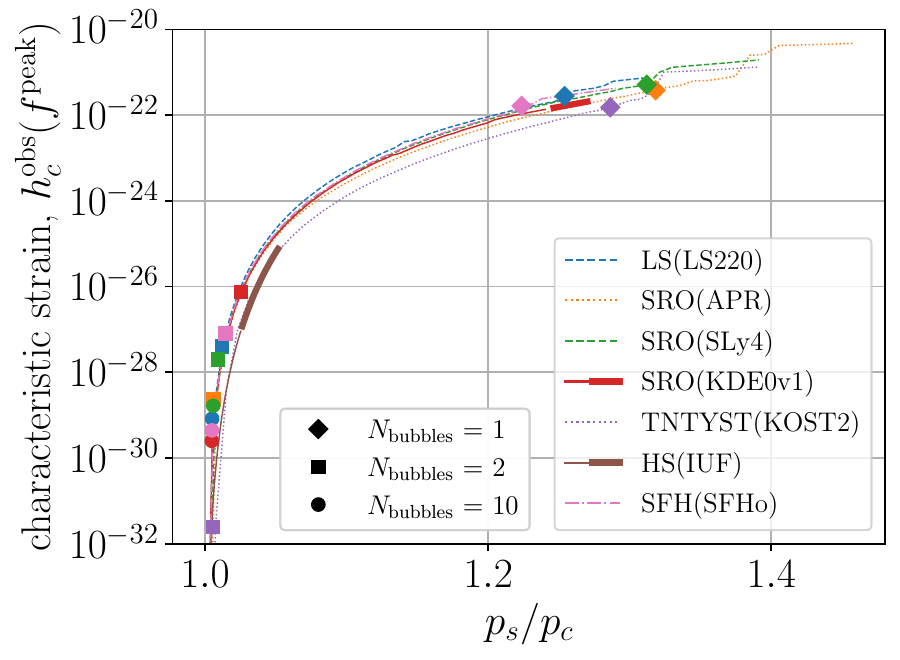}
    \caption{Peak characteristic strain as function of the stalling pressure for a neutron star \SI{8}{kpc} from Earth. The circle, square and diamond markers represent the first time the expected bubble number decreases to 10, 2, and 1. Part of the curves for the SRO(KDE0v1) and HS(IUF) EoS are emphasized to indicate where the conditions $x_q(t_s) > 0.03$ and $N_{\rm bubbles} > 2$ are both fulfilled, implying that the emission of GWs is most likely. $N_{\rm bubbles}$ is always $> 10$ for HS(IUF).}
    \label{fig:h-c}
\end{figure}

\begin{figure}
    \centering
    \includegraphics[width=\columnwidth]{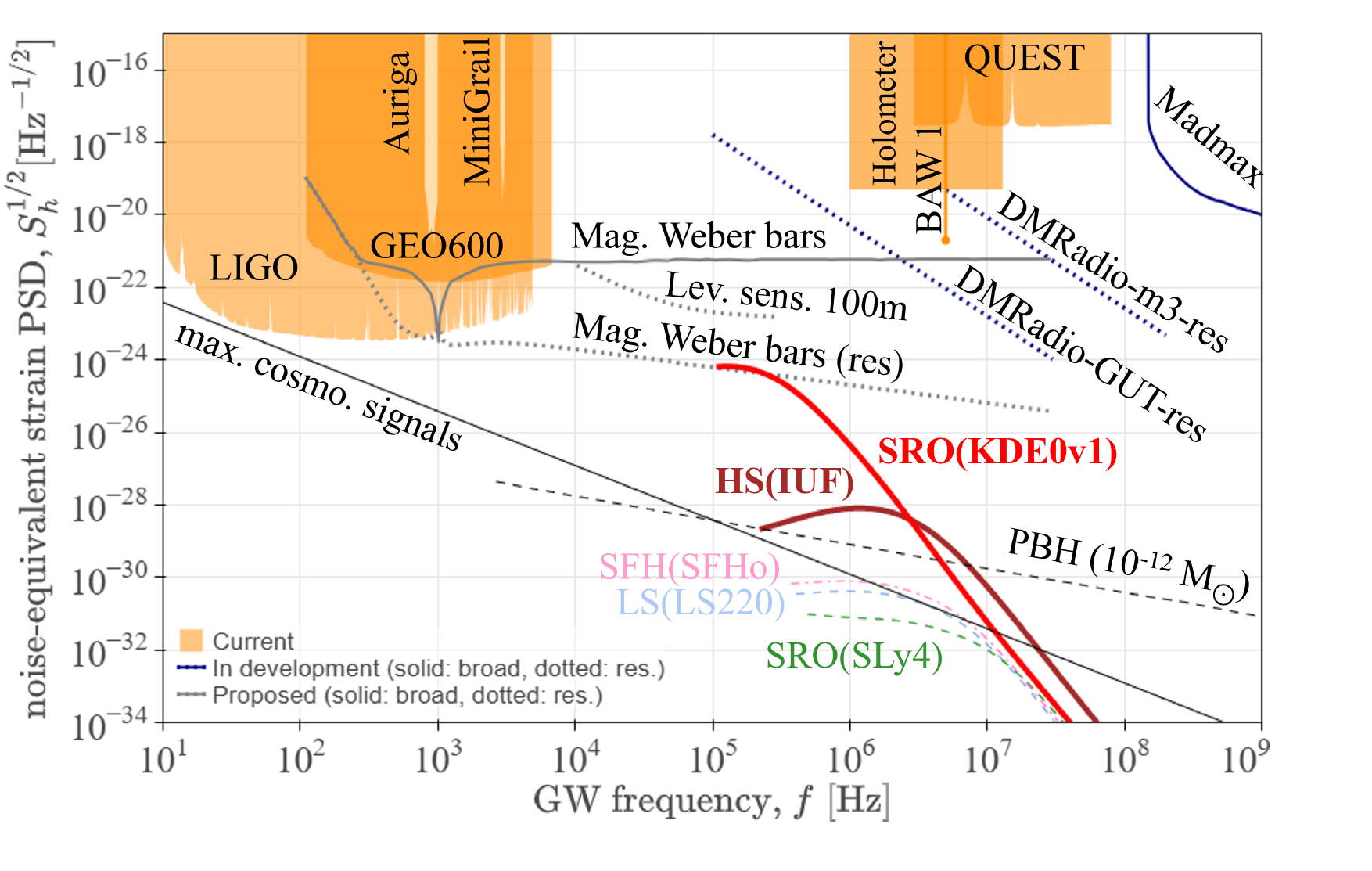}
    \caption{Expected power spectral density (PSD), $\sqrt{S_h} = h_c / \sqrt{f}$, of GWs from a hadron-to-quark phase transition in a supernova at \SI{8}{kpc} (coloured curves) compared to the noise-equivalent strain power spectral density, $S_h^{\rm noise}$, of existing (orange) and proposed (gray) high-frequency GW detectors \cite{Aggarwal:2025noe}. Solid lines indicate broadband detectors, dotted lines those which are only sensitive in a narrow (but tunable) frequency range.
    We show signal predictions for different EoS, in each case choosing $p_s$ such that $N_{\rm bubbles} = 2$. (For HS(IUF), $N_{\rm bubbles}$ is always $ > 2$, so we choose the maximum $p_s$, yielding the maximum GW signal.) The bold curves for HS(IUF) and SRO(KDE0v1) indicate that $x_q(t_s) \gtrsim 0.03$ and $N_{\rm bubbles} \gg 2$ can be satisfied simultaneously, making GW emission likely. Curves are cut off at $f = 1/R_c$. The expected signals for EoS not appearing in the plot are below the plot range.
    The maximum amplitude of cosmological signals and a typical primordial black hole merger signal are shown in black. Figure created using HFGWPlotter \cite{muia_2025_15720342, Aggarwal:2025noe}.}
    \label{fig:gw-spectra}
\end{figure}

\textbf{Results.} We show our main results in \cref{fig:h-c,fig:gw-spectra}. In \cref{fig:h-c} we plot the expected GW characteristic strain at the peak frequency for a source close to the Galactic Center (\SI{8}{kpc} from Earth) as a function of the stalling pressure, $p_s$. The end point of each curve corresponds to the maximum core pressure beyond which the star becomes unstable. At such large $p_s$, it is likely that only a single bubble nucleates, which then expands and eventually stalls without emitting GWs. At lower $p_s$, the number of bubbles is larger, but the final quark matter fraction, $x_q(t_s)$, drops. From the numerical simulations of bubble dynamics detailed in \cref{sec:xq-scaling}, we find that at $x_q(t_s) \lesssim 0.03$, the probability for bubble collisions to set in before the phase transition stalls is $\lesssim 50\%$, implying that in these cases GW emission randomly occurs only in a small fraction of supernovae. As indicated in the plot, only for two of the equations of state considered here -- the HS(IUF) one from refs.~\cite{Hempel:2009mc, Fattoyev:2010mx, Roca-Maza:2008hmq, Hempel:2011mk, Steiner:2012rk} and the SRO(KDE0v1) EoS from refs.~\cite{Lattimer:1991nc, Agrawal:2005ix, Schneider:2017tfi} -- both $N_{\rm bubbles} \geq 2$ and $x_q(t_s) \gtrsim 0.03$ can be satisfied simultaneously. For HS(IUF), the reason is that this EoS is particularly soft around $p_c$ (i.e.\ $dn_{b,h}(p_c)/dp$ is large), implying that small $p_s - p_c$ is sufficient to achieve sizable $x_q$ according to \cref{eq:xq-from-eos}. For SRO(KDE0v1), the reason is the subtle interplay between the growth of $R_c$ and $x_q$ with $p_s$, which implies that the expected number of bubbles first drops as $p_s$ grows, but then rises again, marginally exceeding our somewhat arbitrarily chosen threshold $N_{\rm bubbles} \geq 2$. For this reason, and because the differences between SRO(KDE0v1) and other EoS are marginal, the promising results for SRO(KDE0v1), but also the less promising ones for other EoS, should be taken with a grain of salt. In addition, as bubble nucleation and collision are stochastic processes, \cref{fig:gw-spectra} does not imply that larger signals are impossible for other equations of state -- it only means that the fraction of supernovae that source sizable high-frequency GW signals is $\ll 1$.

In terms of signal amplitude, we find that strains of order $10^{-22}$ may be possible for a galactic source under optimistic assumptions. As can be seen in \cref{fig:gw-spectra}, this is not too far from the reach of proposed experiments. In particular, the resonant magnetic Weber bars proposed in ref.~\cite{Domcke:2024mfu}, corresponding roughly to the DMRadio-GUT setup, look promising.

\textbf{The QCD phase transition inside a supernova is therefore a promising and, together with neutron star mergers \cite{Blas:2022xco}, so far the \emph{only} Standard Model source of potentially detectable high-frequency GWs.}

When a galactic supernova occurs (the rate is 1--2 per century \cite{Rozwadowska:2020nab}), we propose to scour high-frequency GW data recorded within $\mathcal{O}(\SI{10}{sec})$ from the onset of the neutrino signal for a possible stochastic signal lasting $\sim \SIrange{10}{100}{\mu sec}$.
If a signal is found, it will prove the existence of quark matter in neutron star cores, of a first-order phase transition between the two phases, and it will localize this phase transition in the QCD phase diagrams. It will moreover significantly constrain the EoS of both quark matter and hadronic matter. Even the absence of a signal is not without value: given sufficient detector sensitivity, it will exclude certain equations of state that would have predicted a signal.


\textbf{Acknowledgements.}
It is a pleasure to thank Mikel Sanchez-Garitaonandia for useful discussions, and Valerie Domcke for collaboration in the early stages of this project. We thank Mark Hindmarsh and Mika M\"aki for correspondence about {\tt PTtools}.
KB is grateful to CERN for hospitality during several visits which were essential for the completion of this project. KB and JK have been partially supported by the Cluster of Excellence “Precision Physics, Fundamental Interactions, and Structure of Matter” (PRISMA++, EXC 2118/2) funded by the Deutsche Forschungsgemeinschaft (DFG, German Research Foundation) within the German Excellence Strategy (Project No.\ 390831469).
JL was partly supported by the National Research Foundation of Korea (Grant No.RS-2024-00352537), and the CERN--Korea Theoretical Physics Collaboration and Developing
Young High-Energy Theorists (NRF-2012K1A3A2A0105178151).
JvdV was partially supported by the Dutch Research Council (NWO) under project number VI.Veni.212.133.

\clearpage

\begin{center}
    \bfseries APPENDIX
\end{center}
\vspace*{-.7cm}

\appendix
\section{Fluid Dynamics of the Hadron--Quark Phase Transition}
\label{sec:fluid-dynamics}

We now describe how we obtain the bubble wall velocity, $v_w$, the kinetic energy density of the fluid, $\rho_{\rm kin}$, and the fluid velocity profiles used as input for the computation of the gravitational wave spectrum.
We focus here on solutions for bubbles expanding in isolation, which can often be used to predict the GW spectrum of the full collection of bubbles \cite{Hindmarsh:2013xza, Hindmarsh:2015qta, Hindmarsh:2017gnf, Caprini:2019egz}, and which form the necessary input for the sound shell model used in this work.
In contrast to cosmological phase transitions, the phase transition occurring in the core of a proto-neutron star is an \emph{inverse} phase transition, caused by overcompression of the fluid rather than a drop in temperature.
Hydrodynamic solutions for inverse phase transitions were studied in \cite{Barni:2024lkj, Bea:2024bxu, Barni:2025gnm}, and we will follow their approach here.

The equations of motion for the fluid velocity and the enthalpy $w$ follow from energy-momentum conservation,
\begin{align}
    \partial_{\nu}T^{\mu\nu} = 0, \qquad T^{\mu\nu} = w \, u^\mu u^\nu + p \, g^{\mu\nu},
    \label{eq:consTmunu}
\end{align}
with $T^{\mu\nu}$ the energy-momentum tensor of a perfect fluid, $u^\mu = \gamma (1, \vec v)$ the fluid 4-velocity, $p$ the pressure, and $g^{\mu\nu}$ the spacetime metric, which we take to be the Minkowski metric here.
Assuming spherical symmetry, we can project \cref{eq:consTmunu} in the directions parallel to the fluid velocity and perpendicular to it.
This yields a set of differential equations for the fluid velocity in the radial direction, $v$, and the enthalpy, $w$, as a function of the self-similar coordinate $\xi = r / (t - t_0)$, where $r$ is the distance from the bubble center and $t - t_0$ is the time elapsed since the bubble has nucleated.
For details, see e.g.\ refs.~\cite{Espinosa:2010hh, Hindmarsh:2019phv, Barni:2024lkj, Bea:2024bxu}.
Integrating \cref{eq:consTmunu} over an infinitesimal $\xi$ interval around the wall gives us two matching relations for the fluid velocity and equation of state just behind and just in front of the bubble wall.
To close the set of equations and determine the bubble expansion velocity, we need one more condition.
In the absence of a microscopic theory from which we can determine $v_w$, we assume local thermal equilibrium (LTE), which gives us an additional hydrodynamic matching relation \cite{BarrosoMancha:2020fay, Balaji:2020yrx, Ai:2021kak, Ai:2023see}. 
In reality, we expect additional friction sources to be present, modifying the value of the bubble wall velocity. 
Inverse phase transitions have the particular feature that, for certain values of the phase transition strength, no LTE solution exists for the wall velocity~\cite{Barni:2024lkj}. Whenever this happens, we drop the assumption of LTE and choose the largest allowed velocity \emph{below} the would-be LTE solution.

For simplicity, for the purpose of solving the hydrodynamic equations, we will assume that both the quark and the hadron phase are described by pure radiation -- the bag equation of state -- though with different numbers of degrees of freedom, and with a bag constant parameterizing the vacuum energy difference.
In this regime, the hydrodynamic equations simplify significantly. 
We map our realistic hadron equation of state onto the bag equation of state via the value of the phase transition strength \cite{Giese:2020rtr}
\begin{equation}
    \alpha_\theta \equiv \frac{\Delta \theta}{3 (e_h + p_h)} \,.
\end{equation}
Here, $\theta \equiv e - 3p$ denotes the trace of the energy--momentum tensor, and $\Delta\theta \equiv e_q - e_h - 3[p_q - p_h]$ is the change of this quantity across the phase transition.
We extract the energy density, $e$, and pressure density, $p$, directly from the equation of state.
In all of the above quantities, a subscript $h$ refers to the hadron phase, a subscript $q$ to the quark phase.
All quantities are evaluated at the stalling chemical potential $\mu_s$, and the stalling pressure is $p_s = p_h(\mu_s)$.

The kinetic energy of the fluid is 
\begin{align}
    \rho_{\rm kin} &= \frac{3}{v_w^3} \int \! d\xi \, \xi^2 v^2 \gamma^2 (e + p) ,
\end{align}
where all quantities under the integral are understood to be functions of $\xi$.
The root-mean square fluid velocity, used in the estimate of the shock formation time, is given by $\bar U_f = \sqrt{\rho_{\rm kin}/(e_h+p_h)}$.

Finally, we determine $\tilde P_{\rm gw}(z)$ using the sound shell model \cite{Hindmarsh:2016lnk, Hindmarsh:2019phv, RoperPol:2023dzg}, which provides a semi-analytic approach to determining the GW spectrum.
It assumes that the sound waves follow linear hydrodynamics from the moment the bubbles collide. 
As a consequence, the gravitational wave spectrum can be completely determined from the enthalpy and velocity profiles of the single bubble solutions.
We use a slightly modified version of {\tt PTtools} \cite{Hindmarsh_PTtools_2025}, taking our inverse phase transition hydrodynamic profiles as input, to determine the gravitational wave spectrum.

We compute the gravitational wave spectrum {\tt spec} = {\tt pttools.ssmtools.spectrum.SSMSpectrum} using the fluid velocity and enthalpy profiles as input in a custom bubble wrapper. 
To calculate the energy density given by \cref{eq:rho-gw}, we make use of the attribute {\tt spec.spec\_den\_gw} = $\tilde \Omega_{\tt spec\_den\_gw}$ which returns
\begin{align}
    \tilde \Omega_{\tt spec\_den\_gw}
    = 3 \left( \frac{\rho_{\rm kin}}{e} \right)^2 \frac{z^3}{2 \pi^2} \tilde P_{\rm gw}(z) \tau_{\tt PTtools},
\end{align}
where $\tau_{\tt PTtools}$ is the lifetime of the GWs computed by {\tt PTtools}, which we will divide out in order to replace it by the shock formation time $\tau_{\rm sh}$. We can then calculate the energy density as
\begin{align}
    \frac{d \rho_{\rm gw}}{d \log f} = \frac{8\pi G}{3} (e_q + p_q)^2 
    \tilde \Omega_{\tt spec\_den\_gw}
    \frac{\tau_{\rm sh}}{\tau_{\tt PTTools}} r_* \frac{R_c^3}{\tau_{\rm sh}^3} x_q^{4.8}.
\end{align}

\section{Gravitational Wave Signal from Stalled Phase Transitions}
\label{sec:xq-scaling}

To obtain \cref{fig:h-c,fig:gw-spectra} in the main paper, we have computed GW spectra for stalled phase transitions by first using the sound shell model to calculate the signal for a hypothetical complete phase transition, and then rescaling the result accordingly, see \cref{eq:rho-gw}. Here, we empirically determine the corresponding rescaling law, $\rho_\text{gw} \propto \sim x_q^{4.8}$, where $x_q$ is the volume fraction of the stellar core that has transitioned from the hadron phase to the quark phase. We first motivate this scaling qualitatively, and then refine our analytic estimate using lattice simulations of bubble dynamics.

\subsection{Qualitative Arguments}
\label{sec:xq-scaling-qualitative}

The dominant GW production mechanism in the phase transitions of interest to us is from sound waves whose origin are the fluid velocity perturbations around the bubble walls. While bubbles expand, these perturbations grow self-similarly with the bubble, but when bubbles collide and segments of the bubble walls thereby disappear, they decouple and henceforth propagate freely as sound waves. We first consider the evolution of energy density in free sound waves, $\rho_\text{sound}$. In a given time step $dt$, $\rho_\text{sound}$ grows by an amount
\begin{align}
    d\rho_\text{sound} \propto t^2 \times t^3 \times t \times t^2 \, dt .
    \label{eq:drho}
\end{align}
Here,
\begin{itemize}
    \item two powers of $t$ are due to the growth of each bubble during the time interval $dt$, which is proportional to $t^2 dt$.
    
    \item three powers of $t$ come from the total volume of all other bubbles at time $t$. This is the volume that a given ``test bubble'' needs to enter for a collision to begin during the time interval $dt$.
    
    \item one power of $t$ arises from the self-similar scaling of the width of the velocity perturbation around each bubble wall.

    \item two factors of $t$ come from the increasing number of bubbles in the system. The bubble nucleation probability per time interval $dt$ and volume element $dx^3$ is
    \begin{align}
        P_\text{nuc} = \Gamma_0 \exp[\beta (t - t_0)] \, dt \, dx^3 ,
        \label{eq:p-nuc}
    \end{align}
    where $\Gamma_0$ is the nucleation rate at the initial time $t=t_0$, and $\beta$ is the parameter describing the speed of the phase transition. At early times, though, $P_\text{nuc}$ is approximately constant, so the number of bubbles increases linearly with time. The probability that two bubbles collide then scales as $t^2$.
\end{itemize}
Integrating \cref{eq:drho} over time, and taking into account that the GW energy density, $\rho_\text{gw}$ is proportional to $\rho_\text{sound}^2$, we obtain
\begin{align}
    \rho_\text{gw} \propto \rho_\text{sound}^2 \propto t^{18} .
\end{align}
Considering finally that $x_q(t) \propto t^4$, with one power $t$ due to the increase in the number of bubbles and three powers of $t$ coming from the growth of the previously nucleated bubbles, we arrive at the scaling law
\begin{align}
    \rho_\text{gw} \propto x_q^{9/2} .
\end{align}

\subsection{Bubble Dynamics from Simulations}
\label{sec:xq-scaling-simulations}

To corroborate and refine the rough scaling arguments given above, we have simulated bubble nucleation, expansion, and collision on a three-dimensional lattice. We use $80 \times 80 \times 80$ lattice sites in space, and we implement periodic boundary conditions. We simulate up to 100 time steps, with the step size equal to the lattice spacing. In many cases, though, the phase transition completes before the maximum number of time steps has been reached. Each lattice site takes a value of 0 or 1, depending on whether it is still in the hadron phase or has already transitioned to the quark phase.

In each time step, we first determine the number and location of newly nucleated bubbles according to \cref{eq:p-nuc}. We then grow the radii of all bubbles by an amount $v_w dt$, where $v_w$ is the wall velocity. Next, we use the ``marching cubes'' algorithm \cite{Lorensen:1987} in the improved version due to Lewiner \cite{Lewiner:2003} implemented in the \texttt{scikit-image} package \cite{scikit-image} to tesselate the surface of the quark matter region. More precisely, the algorithm (which we have modified to take into account our periodic boundary conditions) turns our 3D lattice data into a triangular mesh describing the surface. The mesh is defined by lists of vertices, edges, and normal vectors. As the normal vectors returned by the marching cubes algorithm are often inaccurate due to discretization errors, we instead keep track of which bubble each vertex belongs to, and we correct the normal vectors using the known locations of the bubble centres.

We are now in a position to compute the energy density contained in the fluid perturbations around the bubble walls, $\rho_\text{fluid}(t)$: each cell of the surface mesh makes a contribution proportional to $dA_i \times r_i$, where $dA_i$ is the area of the $i$-th mesh cell, and $r_i$ is the radius of the bubble to which it belongs. The proportionality to $r_i$ is due to the fact that bubbles -- and their associated fluid perturbations -- expand self-similarly, i.e.\ the thickness of the perturbed fluid shell grows proportional to $r_i$. We can also calculate an estimate $\rho_\text{fluid}^\text{est}(t + dt)$ for the energy density of the fluid perturbations after the \emph{next} time step by shifting all vertices of the mesh by an amount $v_w \, dt$ along the normal direction. $\rho_\text{fluid}^\text{est}(t + dt)$ is, however, an \emph{overestimate} of the real $\rho_\text{fluid}(t + dt)$: it double-counts those mesh cells which intersect between $t$ and $t+dt$. These are precisely the mesh cells whose associated fluid perturbations turn into freely propagating sound waves between $t$ and $t+dt$. In other words, the energy injected into free sound waves between $t$ and $t+dt$ is
\begin{align}
    d\rho_\text{sound} = \rho_\text{fluid}^\text{est}(t + dt)
                       - \rho_\text{fluid}(t + dt).
    \label{eq:drho-sound}
\end{align}
In practical terms, we store the estimate $\rho_\text{fluid}^\text{est}(t + dt)$ until the next time step, where, after having computed $\rho_\text{fluid}(t + dt)$, we can evaluate \cref{eq:drho-sound}. This is the central step of our simulation.

Integrating over all time steps, we thus determine $\rho_\text{sound}(t)$ as a function of time.  We then use $\rho_\text{gw}(t) \propto [\rho_\text{sound}(t)]^2$. We also obtain the volume fraction in the quark phase, $x_q(t)$, directly by counting the lattice cells that have transitioned at any given time. This ultimately allows us to express $\rho_\text{gw}$ as a function of $x_q$.

\begin{figure}
    \centering
    \includegraphics[width=\linewidth]{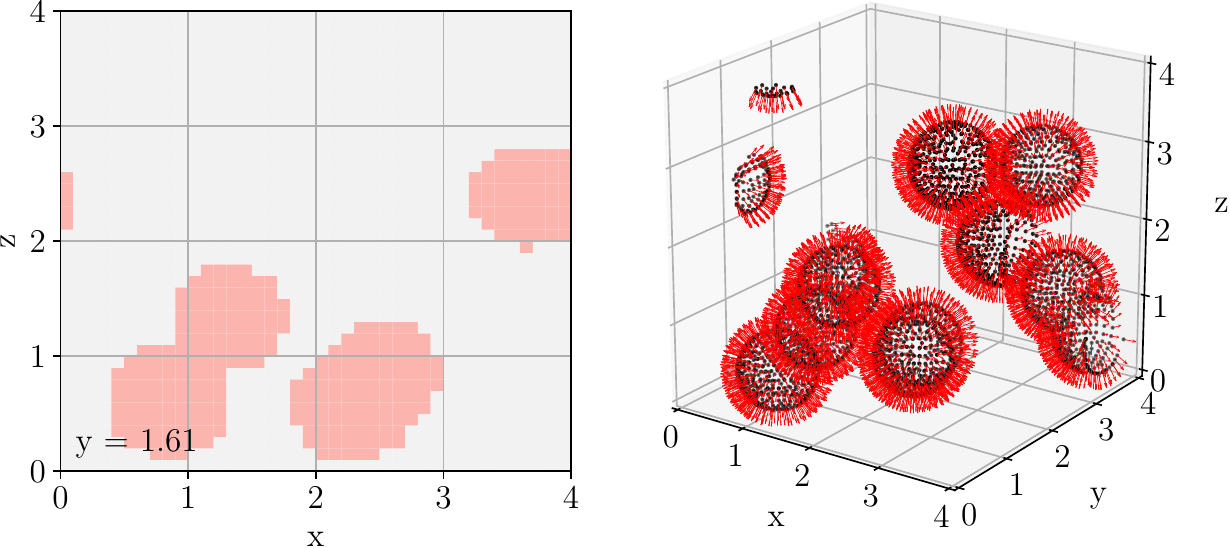}
    \caption{Snapshot from one of our lattice simulations of bubble dynamics, showing on the left a slice through the 3D grid (red regions are in the quark phase, grey regions in the hadron phase), and on the right the corresponding tesselation of the bubble walls, with black dot indicating vertices and red arrows the corresponding normal vector. To improve the readability of this illustration, we have chosen a more coarse-grained grid ($40 \times 40 \times 40$) than for our main simulation, and we have seeded 10 bubbles at random locations by hand rather than nucleating new bubbles in each time step.}
    \label{fig:lattice-simulation}
\end{figure}

\begin{figure}
    \centering
    \includegraphics[width=\linewidth]{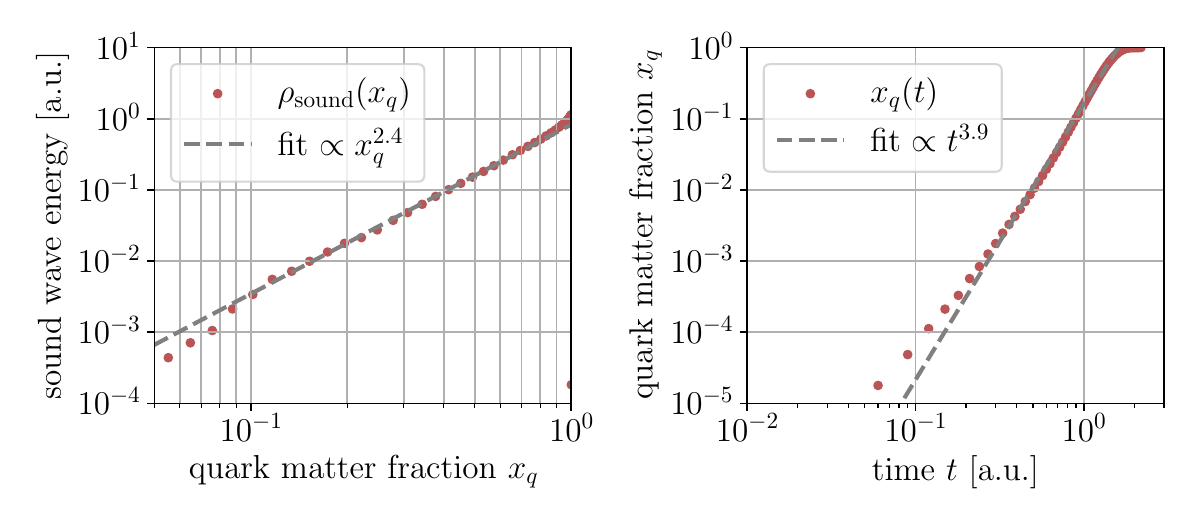}
    \caption{\emph{Left:} energy density in free sound waves, $\rho_\text{sound}$, as a function of the volume fraction in the quark phase, $x_q$. Red dots correspond to our simulation results, while the dashed gray line is a power-law fit. At very early times, the scaling is different due to the very small number of bubbles and due to discretization artefacts. At very late times, the scaling changes again as the shapes of the quark matter regions differ significantly from spherical. 
    \emph{Right:} the scaling of $x_q$ with time, illustrating the expected $x_q(t) \propto t^4$ behaviour. Note that all quantities in this plot are in arbitrary units, which is sufficient since we are only interested in scaling laws.}
    \label{fig:Pv-vs-xq}
\end{figure}

\Cref{fig:lattice-simulation} illustrates our simulation procedure by showing a snapshot at one particular time. The panel on the left shows a slice through the grid data, with red coloured regions corresponding to the quark phase and grey regions to the hadron phase. We can identify five bubbles, one of them wrapping around from $x=4$ to $x=0$ due to the periodic boundary conditions. The other four bubbles are currently undergoing pairwise collisions. The panel on the right shows the tessellation of the bubble walls (black dots) as well as the normal vectors (red arrows) at each vertex. The main result of this section is displayed in \cref{fig:Pv-vs-xq}: the energy density in free sound waves, $\rho_\text{sound}$ as a function of the volume fraction in the quark phase, $x_q$.

We have carried out a number of cross-checks to validate our simulations. In particular:
\begin{itemize}
    \item We have verified that the scaling behaviour shown in \cref{fig:Pv-vs-xq} does not change as we vary the grid spacing.

    \item We have carried out simulations in which bubbles are not nucleated randomly during the simulation according to \cref{eq:p-nuc}, but instead the simulation starts with a fixed number of bubbles nucleated at $t=0$. As expected the scaling behaviour changes in accordance with the expectations based on the qualitative arguments from \cref{sec:xq-scaling-qualitative}. In particular $x_q \propto t^3$ in this case, and $\rho_\text{sound} \propto x_q^{7/3}$.
\end{itemize}

\subsection{Onset of Bubble Collisions}
\label{sec:onset-of-collisions}

Using simulations similar to the ones described above, we also investigate at what quark matter fraction, $x_q$, bubble collisions typically begin. To this end, we seed $N_{\rm bubbles}$ bubbles at random locations, simulate their expansion and coalescence, and follow the evolution of $\rho_{\rm sound}(t)$. Initially, $\rho_{\rm sound}(t)$ is flat close to zero (with small fluctuations due to numerical errors), but eventually it starts rising when the first bubble collisions occur. We find the time, and thereby the $x_q$, at which this onset happens by fitting a broken power law to $\rho_{\rm sound}(t)$ and determining the crossing point of the two segments of the fit.

\begin{figure}
    \centering
    \includegraphics[width=\linewidth]{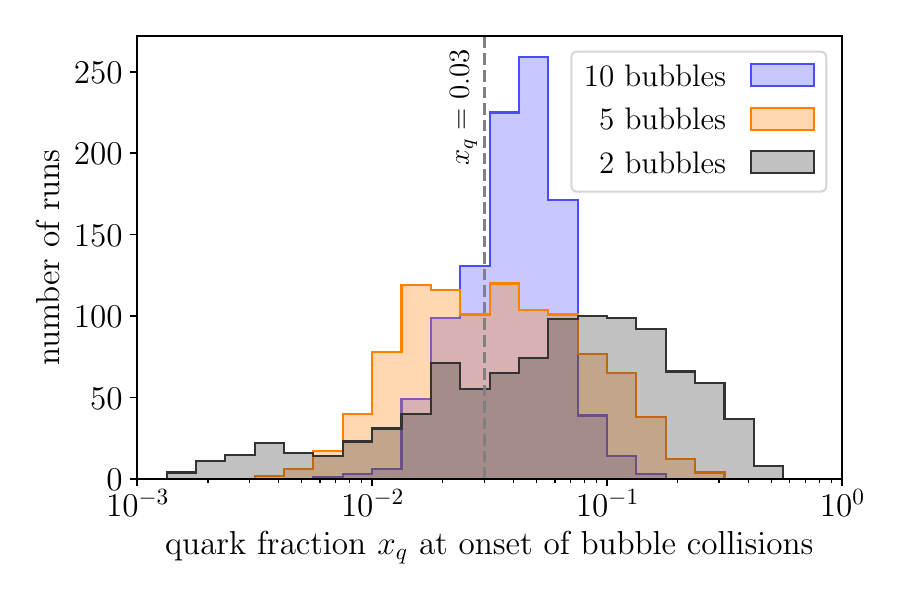}
    \caption{Onset of bubble nucleation in our simulations. For a given number of randomly nucleated seed bubbles, we plot the distribution of the $x_q$ values at which the free sound wave energy, $\rho_{\rm sound}$, begins to increase as bubbles start to collide. We indicate the threshold of $x_q = 0.03$, above which we consider a GW signal reasonably likely.}
    \label{fig:onset-histogram}
\end{figure}

We plot the resulting distribution of onset points in \cref{fig:onset-histogram}. We find that the distribution is largely independent of the number of seed bubbles. This is due to two counteracting effects: on one hand, with more bubbles it is more likely that two of them nucleate close to each other, increasing the probability of collisions. On the other hand, with more bubbles expanding simultaneously, $x_q$ increases more rapidly, so when $x_q$ passes a given threshold, bubbles are still smaller, which reduces the likelihood of collisions.

\Cref{fig:onset-histogram} motivates the threshold $x_q = 0.03$ which we have used in this work to indicate that bubble collisions (and therefore GW emission) is reasonably likely.

\section{Stochastic Signals from Distant Supernovae}
\label{sec:popcorn}

In addition to the potential high-frequency GW signals from Galactic supernovae discussed in this paper, we also expect a background from supernovae in distant galaxies. With a typical signal duration of order $\SI{10}{\mu sec}$ and a supernova rate of order 30 per second in the observable Universe,\footnote{We obtain this estimate using the redshift-dependent supernova rate from ref.~\cite{Ando:2002ky}, see also ref.~\cite{Cocco:2004ac}.} this background will consist of many individual sub-threshold signals with gaps in between. A possible search strategy for such ``popcorn'' signals has been discussed in ref.~\cite{Smith:2017vfk}. Nevertheless, detection will be very challenging: the time-averaged power of the popcorn signal will be some 25 orders of magnitude lower than the signal power during a galactic event, and even the largest ``pops'', coming from supernovae at $\mathcal{O}(\SI{10}{Mpc})$ distances, where the rate reaches $\sim 1$/yr, will be at least six orders of magnitude weaker than a galactic signal.

\bibliographystyle{JHEP}
\bibliography{refs}

\end{document}